\documentclass[aps,prl,twocolumn,groupedaddress,float]{revtex4}

\newcommand{\vdimer}{{\vrule height0.2cm width0.05cm depth0pt}}
\newcommand{\hdimer}{{\hrule height0.05cm width0.2cm depth0pt}}
\newcommand{\vdimers}{{\vrule height0.1cm width0.05cm depth0pt}}
\newcommand{\hdimers}{{\hrule height0.05cm width0.1cm depth0pt}}
\newcommand{\verdimers}{\hbox{\vdimer \hskip 0.1cm \vdimer}}
\newcommand{\hordimers}{\hbox{\vbox{\hdimer \vskip 0.1cm \hdimer}}}
\newcommand{\vdimera}{\hbox{\vdimer \hskip 0.01cm}}
\newcommand{\hdimera}{\hbox{\vbox{\hdimer \vskip 0.05cm }}}
\newcommand{\vdimerb}{\hbox{\vdimers \hskip 0.01cm}}
\newcommand{\hdimerb}{\hbox{\vbox{\hdimers \vskip 0.05cm }}}

\usepackage{graphicx}
\usepackage{amsfonts,amssymb}

\begin{document}

\title{Interacting classical dimers on the square lattice}

\author{Fabien Alet$^1$}
\author{Jesper Lykke Jacobsen$^{1,2}$}
\author{Gr\'egoire Misguich$^1$}
\author{Vincent Pasquier$^1$}
\author{Fr\'ed\'eric Mila$^3$}
\author{Matthias Troyer$^4$}

\affiliation{$^1$Service de Physique Th\'eorique (URA 2306 of CNRS), CEA Saclay, 91191 Gif sur Yvette, France}
\affiliation{$^2$LPTMS, Universit\'e Paris-Sud, B\^atiment 100, 91405 Orsay, France}
\affiliation{$^3$Institute of Theoretical Physics, Ecole Polytechnique F\'ed\'erale de Lausanne, CH-1015 Lausanne, Switzerland}
\affiliation{$^4$Theoretische Physik, ETH Z\"urich, CH-8093 Z\"urich, Switzerland}

\date{\today}

\begin{abstract}

We study a model of close-packed dimers on the square lattice with a
nearest neighbor interaction between parallel dimers. This model corresponds to 
the classical limit of quantum dimer models [D.S. Rokhsar and S.A. Kivelson, Phys.~Rev.~Lett.~{\bf 61}, 2376 (1988)]. 
By means of Monte Carlo and
Transfer Matrix calculations, we show that this system undergoes a
Kosterlitz-Thouless transition separating a low temperature ordered phase
where dimers are aligned in columns from a high temperature critical phase
with continuously varying exponents. This is understood by constructing the
corresponding Coulomb gas, whose coupling constant is computed numerically. 
We also discuss doped models and implications on the finite-temperature phase diagram of quantum dimer models.

\end{abstract}


\maketitle

The model of lattice coverings by hard-core dimers has a long
history~\cite{roberts} in classical statistical physics, culminating in its
solution for arbitrary planar graphs~\cite{kasteleyn}, the calculation of
correlation functions~\cite{fisherstephenson}, and its connection to
Ising~\cite{fisher} or height models~\cite{blote,henley}. With the
introduction of quantum dimer models (QDM) by Rokhsar and
Kivelson~\cite{rk}, and later in connection
with resonating valence bond (RVB) and fractionalization physics in
2D~\cite{QDM} and 3D~\cite{3dq}, dimer models have regained interest.

As a step towards the understanding of the finite-temperature phase diagram of the
QDM, we here study its limit when kinetic terms vanish, which is a classical dimer model with
an interaction that tends to align neighboring dimers. The dimer-dimer correlations which are critical in the noninteracting model~\cite{fisherstephenson} are found to to remain critical, but with continuously varying
exponents, down to a finite temperature where the dimers order through a
Kosterlitz-Thouless (KT) transition~\cite{KT}. Our results are naturally
understood in the Coulomb gas approach~\cite{CG} to the associated height
model~\cite{blote,henley}. Indeed, this interacting dimer model provides one of
the simplest realizations (together with the six-vertex and the XY model) of a
Coulomb gas (CG) with continuously varying exponents. We further study the
effect of doping the model with monomers, and comment on the consequences
for the quantum case.
 
{\it Model }--- We study a model of interacting dimers on
the square lattice, first introduced in the liquid crystal context~\cite{old}, defined by the partition function
\begin{equation}
    \label{eq:ham}
    Z=\sum_{ \{ c \} } \exp \left[-\frac{v}{T}
  ( N^c(\hordimers ) +  N^c(\verdimers )) \right],
\end{equation}
where $T$ is the temperature, the sum runs over all dimer coverings $\{c\}$,
and $N^c(\hordimers)+ N^c(\verdimers)$ is the number of
plaquettes with parallel dimers 
in
configuration $c$. We mainly consider the case $v=-1$, with {\it aligning}
interactions between dimers. Note that the model Eq.~(\ref{eq:ham}) corresponds to the diagonal part of the QDM Hamiltonian~\cite{rk}.
\begin{figure}[tbp]
  \includegraphics[width=7.6cm]{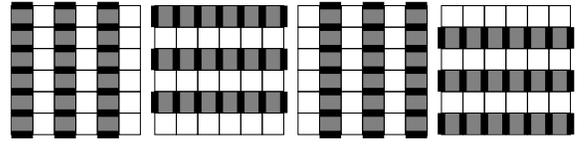}
\caption{The four possible columnar ground states. To each plaquette with a pair of parallel dimers we assign an energy of $+v$ (shaded plaquettes in the figure). }
\label{fig:model}
\end{figure}

At $T=\infty$, Eq.~(\ref{eq:ham}) describes the classical di\-mer
covering problem, and thus displays critical
correlations~\cite{fisherstephenson}. At $T=0$, the dimers order in {\it
columns} and the ground-state is four-fold degenerate, as shown in
Fig.~\ref{fig:model}. This columnar phase breaks
translational and rotational symmetry.

We study Eq.~(\ref{eq:ham}) with Transfer Matrix (TM) techniques
and with a Monte Carlo (MC) directed loop algorithm \cite{sandvik}. The MC
simulations are made on $N=L \times L$ lattices with periodic boundary
conditions (PBC).

\begin{figure}[htbp]
  \includegraphics[width=8cm]{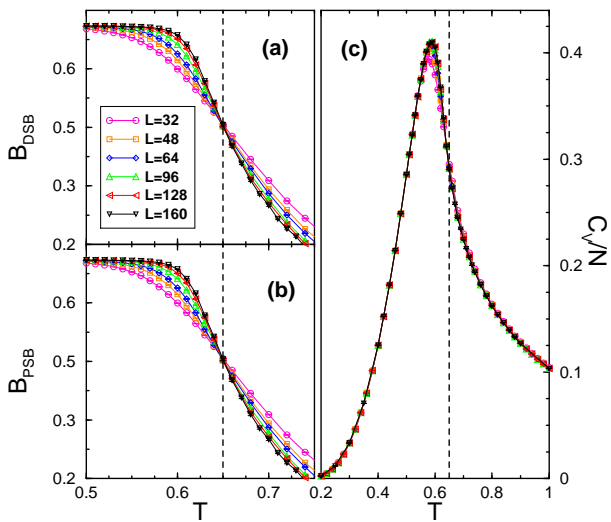}
\caption{(Color online) Dimer symmetry breaking cumulant (a), Pair symmetry breaking cumulant (b), and Specific heat 
per site (c) versus temperature for different system sizes. The dashed lines indicate the estimate
$T_c=0.65(1)$.}
\label{fig:combo}
\end{figure}

{\it Transition to the columnar phase }--- Previous simulations of the QDM on
the square lattice found the existence of a {\it plaquette} phase (which
breaks translational but not $\frac{\pi}{2}$-rotational
symmetry)~\cite{leung}. In the MC simulations of the classical model
Eq.~(\ref{eq:ham}), we also find important plaquette correlations on finite
samples, which nevertheless vanish in the thermodynamic limit~\cite{alet}.
However, they affect the finite-size scaling analysis of the order parameter used in Ref.~\onlinecite{sachdev}. 
For this reason we resort to quantities insensitive to plaquette correlations to detect the entrance into
the columnar phase. Following Ref.~\onlinecite{leung}, we define the dimer
rotational symmetry breaking (DSB) and pair rotational symmetry breaking (PSB)
order parameters as:
\begin{eqnarray}
\label{eq:dsb}
{\rm DSB} &=& N^{-1} \left| N^c( \hdimera ) -  N^c(\vdimera ) \right|
 \nonumber \\
{\rm PSB} &=& N^{-1} \left| N^c ( \hordimers ) -  N^c(\verdimers ) \right|, 
\end{eqnarray}
with $N^c(\hdimera )$ ($N^c(\vdimera)$) the number of horizontal
(vertical) dimers in the configuration $c$. Both quantities show
long-range order in a columnar phase, but vanish in a plaquette phase. To
locate the transition with high precision, we study Binder
cumulants~\cite{binder}, defined as
$B_{\rm DSB}=1-\langle {\rm DSB}^4\rangle / (3\langle {\rm DSB}^2\rangle^2)$
and 
$B_{\rm PSB}=1-\langle {\rm PSB}^4\rangle / (3\langle {\rm PSB}^2\rangle^2)$;
here $\langle \cdots \rangle$ denotes the ensemble average. In the
thermodynamic limit, these cumulants saturate to $2/3$ in a long-range ordered
phase, and vanish in a disordered phase. For both cumulants, the
curves for different $L$ all cross at a unique temperature $T$
(see Figs.~\ref{fig:combo}(a) and (b)), which we identify
as the transition temperature $T_{\rm c}=0.65(1)$ to the columnar phase.

The specific heat per site $C_v/N$ displays a peak, which does
{\it not} diverge in the thermodynamic limit [Fig.~\ref{fig:combo}(c)].
This points towards a second order transition with a critical exponent
$\alpha\leq 0$, or a KT transition. The peak of $C_v/N$ is
located slightly below the value of $T_c$ determined by the cumulants.
A featureless
specific heat at $T_c$ is a strong indication of a KT
transition.

{\it High temperature phase }--- For $T>T_c$ we calculate the dimer-dimer
correlation function
\begin{equation}
\label{eq:dd}
 G({\bf x})= \langle n_{\hdimerb}({\bf r})n_{\hdimerb}({\bf r+x}) \rangle
 -1/16, 
\end{equation}
where $n_{\hdimerb}({\bf r})=1$ for a horizontal dimer at site {\bf r}, and
$0$ otherwise. The constant $1/16$ stands for the dimer density squared.
Taking ${\bf x}=(x,0)$ we find
that $G(x)$ is staggered with $x$ at any $T>T_c$. We also
calculate the monomer-monomer correlation function $M({\bf x})$, 
proportional to the number of configurations with two test monomers
separated by ${\bf x}$, and normalized by $M(1)=1$~\cite{fisherstephenson,krauth,sandvik}. 
$M({\bf x})$ can be computed in the loop building process of the MC simulation~\cite{sandvik}. 
For a bipartite system, $M({\bf x})=0$ for monomers on the same sublattice.

\begin{figure}[tbp]
  \includegraphics[width=8cm]{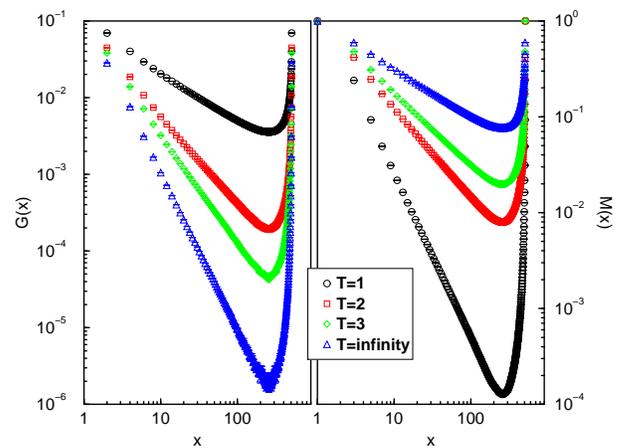}
\caption{(Color online) Dimer-Dimer $G(x)$ (left) and Monomer-Monomer $M(x)$ (right) 
correlation functions versus distance $x$ in a log-log scale for various temperatures.}
\label{fig:Correl}
\end{figure}

Anticipating the results, we look for algebraic decay of these correlators for $T>T_c$ 
and define exponents $\alpha_d$ and $\alpha_m$ by
$G(x)\sim (-)^x x^{-\alpha_d}$ and $M(x)\sim x^{-\alpha_m}$ for large $x$. At
$T=\infty$, we have the exact results $\alpha_d=2$ and
$\alpha_m=1/2$~\cite{fisherstephenson}. In Fig.~\ref{fig:Correl}, we show $G(x)$ and $M(x)$ calculated on an $L=512$
sample for various temperatures $T>T_{\rm c}$. Both correlators indeed show
{\it power-law decay} for all $T>T_{\rm c}$, and the exponents $\alpha_d$ and
$\alpha_m$ appear to {\it vary continuously} with $T$. The power-law decay is
eventually truncated around $L/2$ due to the PBC, making a high precision fit for $\alpha_d(T)$ and $\alpha_m(T)$ difficult.

We therefore turn to TM calculations, which are performed on semi-infinite
cylinders of {\em even} circumference $L$. By convention, the space coordinate
$x$ is horizontal and time $t$ runs upwards. 
Placing arrows on all lattice edges, such that even (odd) sites are sources (sinks) of
four arrows, we define a (staggered) reference configuration $c_0$ by placing dimers on
up-pointing arrows only. Superposing a given configuration $c$ with $c_0$
gives a transition graph with a conserved number $W_x$ of (oriented) time-like strings.
More precisely, $W_x=\sum_x \big( n^\uparrow_{\vdimerb}(x,t)- n^\downarrow_{\vdimerb}(x,t) \big)
\in \{-\frac{L}{2},\ldots,\frac{L}{2} \}$ is independent of $t$ and labels the
blocks of the TM. The block $W_x \neq 0$ corresponds to a defect of $|W_x|$
monomers on the same sublattice at $t=-\infty$, and $|W_x|$ monomers on the
opposite sublattice at $t=\infty$. This can be seen by taking a $W_x=0$ configuration and 
shifting its dimers along some time-like string: this changes $W_x$ by $\pm 1$ and introduces 
a monomer at either end of the string.

Ordering the TM eigenvalues as $\lambda^{W_x}_1 \ge \lambda^{W_x}_2 \ge \ldots$,
we expect $\lambda^0_1$ (no monomers) to dominate all other eigenvalues.
Using standard predictions of conformal field theory~\cite{cft}, the
$L$-dependence of $\lambda^0_1$ gives the central charge $c$;
the ratio $\lambda^0_1/\lambda^0_2$ determines $\alpha_d$
and $\lambda^0_1/\lambda^1_1$ gives $\alpha_m$. Note that the eigenvectors
corresponding to $\lambda^0_1$ (resp.~$\lambda^0_2$) are even (resp.~odd)
under the shift $x\to x+1$.

To account for the interactions in Eq.~(\ref{eq:ham}), the TM states
must encode vertical {\em and} horizontal dimer occupancies within a row
\cite{alet}.
We use sparse-matrix factorization, and the state space is built from a
reference state in the given $W_x$-sector by means of hashing techniques.
Computations were performed up to $L=18$, with 1\,728\,292 states in the
$W_x=0$ sector.

\begin{figure}[tbp]
  \includegraphics[width=8cm]{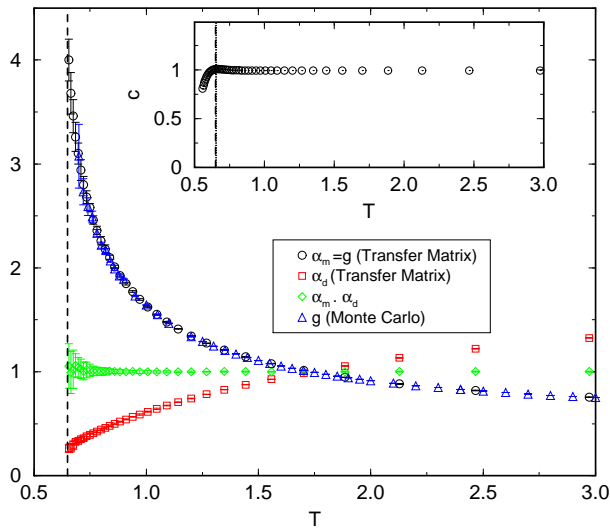}
\caption{(Color online) Dimer-dimer $\alpha_d$ and monomer-monomer $\alpha_m$ exponents (and their product) obtained by 
TM calculations, and coupling constant $g$ of the CG as obtained by MC, all versus temperature $T$. Inset: central charge $c$ versus $T$. 
The dashed lines denote the MC estimate of $T_c$.}
\label{fig:c}
\end{figure}

In Fig.~\ref{fig:c}, we show $\alpha_m$ and $\alpha_d$ as functions of
temperature (error bars are determined from the finite size scaling analysis). $\alpha_m$ increases from $1/2$ at $T=\infty$ to $4.0(2)$ at the
transition temperature $T_{\rm c}$, and then diverges below $T_{\rm c}$ (not shown in Fig.~\ref{fig:c}).
$\alpha_d$ decreases from $2$ at $T=\infty$ to $0.26(4)$ at $T_{\rm c}$, and
vanishes below (not shown). We also plot in Fig.~\ref{fig:c} the product $\alpha_m \cdot
\alpha_d$, which equals $1$ within error bars for {\it any} $T \ge T_{\rm c}$.  The continuous variation of
both $\alpha_m$ and $\alpha_d$ as functions of $T$ indicates criticality for
$T_{\rm c} \le T \le \infty$. This is corroborated by the central charge $c$
as a function of $T$ (see inset of Fig.~\ref{fig:c}), which is $1.00(1)$ in
the critical phase and drops abruptly below $T_{\rm c}$. $T_{\rm c}$ can be estimated
from the TM by $\alpha_m(T_{\rm c})=4$, the jump in $c$, or $\alpha_d(T_{\rm
c})=1/4$. The exact values quoted are accounted for by the CG formalism discussed below. 
The first determination is the most precise and gives $T_{\rm
c}=0.65(1)$, in full agreement with the MC results.

{\it Coulomb gas representation }--- We now account for the above findings
(KT transition, critical phase with continuously varying
exponents) by dressing the model's CG representation~\cite{CG}.
Close-packed dimers on the square lattice admit a height
representation~\cite{blote,henley}, where a height $z$ is assigned (up to an
overall constant) to each plaquette in the following way: going
counterclockwise around a site on the even sublattice, $z$ changes by $+3/4$ when crossing a dimer 
and by $-1/4$ across an empty bond. In these units, a monomer corresponds to a
dislocation of $1$ in the height. To describe the long-wavelength physics of
our model, we follow Ref.~\onlinecite{henley} and coarse-grain the local height
variable $z$ to a continuum field $h({\bf r})$, for which we postulate the
following action:
\begin{equation}
 S= \int d{\bf r} \, \left[ \pi g |\nabla h({\bf r})|^2 +
 V \cos \big(2\pi p h({\bf r})\big) \right]
\label{eq:eff}
\end{equation}
The first term involving the CG coupling $g$~\cite{note.g} accounts for the
dimer entropy associated with a given coarse-grained height $h({\bf r})$; it
dominates in the high-$T$ {\em rough} phase. The second term is a locking
potential of strength $V>0$ (the actual value of $V$ is immaterial) that
favors dimer ordering in one of the $p=4$ ``ideal'' states \cite{henley}
(see Fig.~\ref{fig:model}) satisfying $h({\bf r})=1/8$, $3/8$, $5/8$ and $7/8
\mbox{ (mod } 1)$; it dominates in the low-$T$ {\em flat} phase. When $T
\rightarrow T_c$, the locking potential becomes irrelevant and the height model
undergoes a roughening transition.

At $T>T_c$ the height field is thus free and described by a Coulomb gas
with electric charges $e$ only \cite{CG}.
Dual, magnetic charges $m$ correspond to height dislocations, which can be
inserted ``by hand'' through monomers or appropriate boundary conditions (cf.~the above
discussion of the TM). Units are such that $e$ and $m$ are integers. The
exponent of an electromagnetic operator of charge $(e,m)$ reads
$\alpha(e,m)=g^{-1} e^2 + g m^2$ \cite{CG}. We identify the leading
dimer and monomer correlations by $\alpha_d = \alpha(1,0)$ and
$\alpha_m = \alpha(0,1)$. The exact results of Ref.~\onlinecite{fisherstephenson}
then fix $g=1/2$ at $T=\infty$ \cite{henley};
in general $g=g(T)$ must
depend on $T$ to account for the continuously varying exponents.

More generally, electric charges define the vertex operators $V_e({\bf r}) =
\ :\!\! \exp \big(2\pi i e h({\bf r}) \big) \!\! :$~appearing in the Fourier expansion of
any operator which is periodic in $h({\bf r})$. Note that $V_1$
appears as a continuum limit contribution to the dimer operator \cite{fradkin}.
 The locking potential corresponds to $V_4$ whose marginality determines
$g(T_{\rm c})=4$.

To check these CG predictions, we calculate $g(T)$ from the TM results
(see Fig.~\ref{fig:c}), using $\alpha_m=g$. Alternatively, $g(T)$ is
accessible from MC calculations of the
fluctuations of the {\it winding number} $\langle W^2 \rangle =\frac12 \langle
W_x^2+W_y^2 \rangle$. The winding number $W_x$ (resp.~$W_y$) is the local
height difference accumulated by a path winding around the torus with PBC in
the $x$ (resp.~$y$) direction.
$\langle W^2 \rangle$ is simply related to $g$ in the free-field point of view ~\cite{free}.
By separating the average tilt from the spatial fluctuations of the height, the partition function
is written as a sum over $W_x$  and $W_y$. The associated Boltzmann weight 
is given by the action of Eq.~(\ref{eq:eff}) evaluated with its classical solution $(\Delta h({\bf r})=0)$ 
given by a linear height $h(x,y)=\frac{x}{L_x}W_x+\frac{y}{L_y}W_y$
(and $V=0$). Then
\begin{equation}
\langle W^2\rangle= \sum_{n \in \mathbb{Z}}n^2 {\rm e}^{-g\pi
n^2}  / \sum_{n \in \mathbb{Z}}{\rm e}^{-g\pi n^2} .
\label{eq:g2W}
\end{equation}
Note that at $g=1/2$, this is a non-trivial statement about the Pfaffian formulation of the partition function~\cite{kasteleyn}.
In Fig.~\ref{fig:c}, we show the MC results for $g$ obtained from $\langle W^2\rangle$ and Eq.~(\ref{eq:g2W}), which match 
with high precision the TM evaluation. This validates the CG scenario for the transition in our model.

The CG description also holds for $v=+1$, i.e., when dimer alignment
is penalized. In this case, the CG coupling decreases from $g=1/2$ at
$T=\infty$ to $g=0$ at $T=T_t$. As $T \rightarrow T_t$ the height model
becomes unstable and undergoes a tilting transition. When $T<T_t$ the
winding $\langle W^2 \rangle$ becomes extensive. Analysis of the
$L$-dependence of the effective central charge gives the estimate
$T_t=0.449(1)$. More details will appear in Ref.~\onlinecite{alet}.

{\it Doping with monomers }--- If we allow monomers in the model, the height mapping is lost but the CG description stays valid.
For $T>T_c$ introducing monomers is a relevant perturbation ($\alpha_m < 4$) and destroys the critical phase, but it becomes marginal at $T_c$. Denoting the weight of a monomer by ${\rm e}^{\mu/T}$ and the
chemical potential by $\mu$, we therefore expect a further critical
line $\mu(T)$ to emanate from $\mu(T_c)=-\infty$ and extend towards
decreasing $T$. Physically this line separates a (weakly) doped dimer crystal from a liquid of dimers and
monomers. The ground-state has only monomers for $\mu/|v|>1/2$ but is a 4-fold degenerate
monomer-free crystal when $\mu/|v|<1/2$. The transition is thus first order at $T=0$
and presumably remains so for small $T$ as both phases have {\it gapped} elementary excitations at $\mu/|v|=1/2$.
However, the transition at low monomer doping is second order, so it must cease to be continuous
at some finite $(T_*,\mu_*)$. Using a modified TM we estimate
$T_* = 0.39(4)$ and $\mu_*/|v|=0.25(10)$. Along the line,
for $T_* < T < T_c$, the TM results are consistent with $c=1$ and
a $T$-independent exponent $\alpha_d=1/4$. Less relevant operators
however have $T$-dependent exponents.

{\it Discussion and conclusion }--- In conclusion, we have studied a model of dimers on the square lattice
with an interaction that favors dimer alignment. We find that this model undergoes a 
KT transition separating a high-$T$ critical phase with continuously varying exponents from a low-$T$ 
crystalline phase. The transition is understood as the roughening transition in the height representation or equivalently as a 
proliferation of $e=4$ electric charges in the CG framework. Our model
captures naturally all the contents of the CG: electric (dimers) and magnetic (monomers) charges, and a $T$-dependent coupling 
constant $g$ driving the transition. Concerning quantum systems, the rough phase should survive in a large part of the finite-temperature phase 
diagram of the QDM (above the melting temperatures of the ordered phases) and especially so in the vicinity of the Rokhsar-Kivelson 
point~\cite{rk}. We thus expect continuously varying exponents parametrized by a CG coupling constant
at sufficiently high $T$. Our results also indicate that doping the QDM would immediately destroy this critical phase.
Detailed finite-temperature properties of the QDM yet remain to be explored. It would be interesting as well to study quantum Hamiltonians with ground state corresponding to Eq.~(\ref{eq:ham})~\cite{grk}.

{\it Acknowledgments }--- We thank C. L. Henley, W. Krauth, R. Moessner and A. Ralko for fruitful discussions. The MC calculations were performed on the Gallega cluster at SPhT using the ALPS libraries~\cite{ALPS}.


\begin{thebibliography}{99}
\bibitem{note.g} We use here the notations of Ref.~\onlinecite{CG}. $g$ is related to the stiffness constant $K$ used in Refs.~\onlinecite{henley} by $g=8K/\pi$.
\end{thebibliography}
\end{document}